\documentclass[final,3p,times,twocolumn]{elsarticle}

\usepackage{graphicx}

\usepackage{amsmath,amssymb}

\def\tref#1#2#3{{#1} (#2) #3}
\newcommand{\ii}{\ensuremath{\mathrm{i}}}
\newcommand{\e}{\ensuremath{\mathrm{e}}}
\newcommand{\ppi}{\ensuremath{\mathrm{\pi}}}
\newcommand{\dif}[1]{\ensuremath{ \mathrm{d}\,#1 }}

\newcommand{\alpi}{\ensuremath{\frac{\alpha}{\pi}}}

\biboptions{comma,sort&compress}

\journal{Nuc. Phys. (Proc. Suppl.)}

\begin{document}

\begin{frontmatter}



\title{Chiral-Symmetry-Violating Effects and Near-Maximal Mixing of Scalar Gluonium and Quark Mesons}


\author[label1]{T.G. Steele\corref{cor1}}
\address[label1]{Department of Physics and Engineering Physics, University of Saskatchewan, Saskatoon, SK, S7N 5E2, Canada}
\cortext[cor1]{Speaker}

\author[label2]{D. Harnett}
\address[label2]{Department of Physics, University of the Fraser Valley, Abbotsford, BC, V2S 7M8, Canada}

\author[label1]{R.T. Kleiv}

\author[label3]{K. Moats}
\address[label3]{Department of Physics, Carleton University, Ottawa, ON, K1S 5B6, Canada}

\begin{abstract}
Gaussian QCD sum-rules are used to analyze all possible two-point correlation functions of scalar gluonic and quark currents. The independent predictions of the masses and relative coupling strengths from the different correlators are remarkably consistent with a scenario of two  scalar states that couple to nearly-maximal mixtures of quark and gluonic currents.
\end{abstract}


\end{frontmatter}


One of the most fascinating conceptual predictions of QCD is the possibility of exotic hadronic states composed purely of gluons.  Such glueball states can have the same $J^{PC}$ quantum numbers as conventional quark ($q\bar q$) mesons, and could thus result in supernumerary states beyond what is expected from $q\bar q$ structures alone.  The existence of nineteen scalar $J^{PC}=0^{++}$ mesons below $2\,{\rm GeV}$ \cite{pdg} is suggestive of an  isoscalar glueball in addition to a $q\bar q$ nonet and a $q\bar qq\bar q$ nonet.   However, the observed isoscalars can be mixtures of glueballs and $q\bar q$ mesons,  which could obscure unique phenomenological signatures of glueballs (e.g., suppressed $\gamma\gamma$ decays).  It is thus important to determine how the glueball and $q\bar q$ content is distributed amongst the observed scalar mesons.

In this paper, we briefly review the Gaussian QCD sum-rule analysis of the mixing  of scalar gluonium and quark mesons \cite{Harnett:2008cw,diagonal_GSRs}.  Although our findings are complementary to QCD Laplace sum-rules \cite{laplace_glue} (see Ref.~\cite{narison_review} for a comprehensive review) that conclude that an approximately
$1.6\,{\rm GeV}$ state is a glueball-$q\bar q$ mixture,   a unique aspect of our approach is a comprehensive combined analysis of all possible two-point correlation functions that include glueball and $q\bar q$ currents.\footnote{Key findings from chiral Lagrangians, lattice QCD,  and other theoretical approaches provide further support for the scenario of a $1.5\,{\rm GeV}$ state with  significant gluonium content \cite{other_stuff,lattice}. } Important insights that emerge from our analysis include the crucial role of chiral violating effects from QCD condensates and instantons for a consistent mixing scenario.
 
Within a QCD sum-rules approach, the
composition of a hadronic state may be identified through its coupling
to a field-theoretic current. 
 In particular, we use an OZI-inspired definition that $q\bar q$ states $\vert q\rangle$ couple to the light quark current $J_q=m_q\bar q q$ but decouple from the gluonic current $J_G=\alpha G^2=\alpha G^a_{\mu\nu}G^a_{\mu\nu}$
\begin{equation}
\langle 0\vert J_q \vert q\rangle \ne 0 ~,~  \langle 0\vert J_G \vert q\rangle \approx 0\,.
\end{equation}
Conversely, glueball states  $\vert G\rangle$  couple to  the gluonic current  but decouple from   the
(light) quark  current 
\begin{equation}
\langle 0\vert J_q \vert G\rangle \approx 0 ~,~  \langle 0\vert J_G \vert G\rangle \ne  0\,.
\end{equation}
However, mixed states $\vert M\rangle$ couple to both gluonic and quark currents
\begin{equation}
\langle 0\vert J_q \vert M\rangle \ne 0 ~,~  \langle 0\vert J_G \vert M\rangle \ne 0 \,.
\end{equation}
Three possible correlation functions can then be formed from these currents ($Q^2\equiv -q^2$):
\begin{gather}
  \Pi_{GG}\left(Q^2\right)=\ii\int\,\mathrm{d}^4x\,\e^{\ii q\cdot x}\langle 0
  \vert T\left[ J_G(x)J_G(0)\right] \vert 0 \rangle
  \label{diag_glue}
\\
    \Pi_{qq}\left(Q^2\right) = \ii\,\int\,\mathrm{d}^4x\;\e^{\ii q\cdot x}
  \langle 0 | T \left[J_q (x) J_q (0)\right] |0\rangle
  \label{diag_quark}
\\
  \Pi_{Gq}\left(Q^2\right)=\ii\int\,\mathrm{d}^4x\,\e^{\ii q\cdot x}\langle 0
  \vert T\left[ J_G(x)J_q(0)\right] \vert 0 \rangle\,.
  \label{non_diag}
 \end{gather}
In the pure (un-mixed) states scenario there is no intermediate state that couples to both currents, the non-diagonal correlator 
\eqref{non_diag} will be suppressed, and in the absence of an accidental degeneracy, we expect signals of different states from the diagonal gluonic correlation function \eqref{diag_glue} and the diagonal quark-current correlator \eqref{diag_quark}.  However, if the states are mixtures, then no suppression of  non-diagonal correlator \eqref{non_diag} occurs, and signals of the same states must occur in all three correlators.  Thus the key signature of mixed glueball-$q\bar q$ states is  a substantial (i.e., not suppressed) non-diagonal correlator. 

The original analysis of the non-diagonal correlator \cite{paver} has been extended to include  contributions of the chiral violating mixed condensate and instantons \cite{Harnett:2008cw}.  The calculation is quite subtle; one needs to take into account the renormalization of the gluonic current\footnote{In this notation, 
$R$ denotes a renormalized composite
operator, $B$ denotes  bare
quantities, and 
our convention for dimensional regularization uses $D=4+2\epsilon$ spacetime dimensions. }
\begin{equation}
G_R^2=  \left(1+\frac{\beta_0}{\epsilon}\frac{\alpha}{\pi} \right) G^2_B
-4\frac{\alpha}{\pi}\frac{1}{\epsilon}\left(m_u\overline{u}u + m_d\overline{d}d\right)_B
+\ldots
\end{equation}
and one must also take into account the mixing of condensate coefficients in the operator-product expansion \cite{OPE_mix}.  The  result for the non-diagonal correlator is \cite{Harnett:2008cw}
\begin{gather}
\begin{split}
\Pi_{gq}\left(Q^2\right)=&m_q^2Q^2\left[A_0L+A_1L^2  \right]
+ m_q\langle \bar qq\rangle C_0 L 
\\&
+m_q^2\langle \alpha G^2\rangle \frac{1}{Q^2}\left[B_0+B_1 L 
\right]
+m_q\left\langle \bar q \sigma G q\right\rangle\frac{D_0}{Q^2}
\\&-
8\sqrt{3 n_c} m_q \rho Q^2 \sqrt{\rho^2Q^2} K_1\left(\rho Q\right) K_2\left(\rho Q\right)
\end{split}
\label{final_corr}
\\
\begin{split}
A_0=-\frac{23}{2\pi}\left(\alpi\right)^2\,,~A_1=\frac{3}{2\pi}\left(\alpi\right)^2\,,~
C_0=-8\pi\left(\frac{\alpha}{\pi}\right)^2
\\D_0=4\alpha
\,,
B_0=6\frac{\alpha}{\pi}\,,~B_1=-2\frac{\alpha}{\pi}\,,~ L=\log\left[\frac{Q^2}{\nu^2}\right] \,.
\end{split}
\end{gather}
where the dilute instanton liquid model \cite{DIL} has been employed.
Eq.~\eqref{final_corr} demonstrates that perturbative and gluon condensate contributions have a greater chiral-suppression compared with the chiral-violating effects of the quark condensate, mixed condensate, and instanton.  Using the definition 
\begin{equation}
\langle 0\vert J_G\vert M\rangle=f_G\,,~\langle 0\vert J_q\vert M\rangle=f_q
\end{equation}
for the coupling of a mixed state to the currents, the (dominant) leading-order perturbative contributions to the diagonal correlation functions (\ref{diag_glue},\ref{diag_quark}) give \cite{Harnett:2008cw}

\begin{equation}
 f^2_G\sim \left(\frac{\alpha}{\pi}\right)^2 E^4~,~f^2_q\sim m_q^2 E^2~
\end{equation}
where $E$ is a typical hadronic sum-rule scale.  Combined with \eqref{final_corr}, the approximate contributions to the mixing angle $\theta$ from perturbative and gluon-condensate effects are chirally-suppressed
\begin{equation}
 \sin{2\theta}\sim \frac{\alpha}{\pi}\frac{m_q}{E}\ll 1\,,\sin{2\theta}\sim \frac{m_q}{E}\frac{ \langle\alpha G^2\rangle}{E^4}\ll 1\,.
\end{equation}
However, the quark and mixed-condensate contributions to the mixing angle do not experience  chiral suppression\footnote{Unfortunately, a simple scaling-argument estimate for the instanton is not possible.}
 \begin{equation}
  \sin{2\theta}\sim \frac{\alpha}{\pi} \frac{\left\langle \bar q q\right \rangle}{E^3}\,,\sin{2\theta}\sim \frac{\left\langle \bar q \sigma Gq\right\rangle}{E^5}=\frac{M_0^2}{E^2}\frac{\left\langle \bar q q\right \rangle}{E^3}
 \end{equation}
because the condensate scales $\langle \bar q q\rangle$ and $M_0$ [see Eq.~\eqref{mixed_cond_value}] are comparable to the hadronic scale $E$.  In fact, one can see that the mixed condensate, absent in the original analysis \cite{paver}, is more important than the quark condensate.   
We thus conclude that chiral-violating effects can lead to a significant glueball-$q\bar q$ mixing angle, implying that mixing of gluonic and $q\bar q$ degrees of freedom has a non-perturbative origin.   Qualitatively, this conclusion is similar to that obtained for glueball decays \cite{hongying} and to that of Ref.~\cite{kochelev}  which  demonstrated that instantons can lead to a significant mixing between glueballs and (heavy quark) mesons in the pseudoscalar channel.

To go
beyond order-of-magnitude estimates of the mixing angle, a sum-rule
methodology that is sensitive to multiple states is desirable. Gaussian
sum-rules (GSRs) are ideal for this purpose. These have the form
\cite{gauss}
\begin{equation}
G_0\left(\hat s,\tau\right)=\frac{1}{\sqrt{4\ppi\tau}} \int\limits_{t_0}^\infty
\exp\left[\frac{-\left(t-\hat{s}\right)^2}{4\tau}\right]\,\frac{1}{\ppi}\rho(t)\;\dif{t}
\label{basic_gauss}
\end{equation}
and relate a QCD calculation $G_0\left(\hat s,\tau\right)$ to  its associated hadronic spectral function
$\rho(t)$ with threshold $t_0$.  Although  the quantity $\tau$ (corresponding to the duality interval) is constrained by QCD, the peak $\hat s$ of the Gaussian kernel is unconstrained.  
This implies that as $\hat s$ is varied, the resonance peaks of the
spectral function are probed with equal sensitivity, thereby permitting
multiple states to be resolved. 
This behaviour should be contrasted with Laplace sum-rules which exponentially-suppress excited states, thereby enhancing the ground state.  

The original development of GSRs used the heat-evolution equation to demonstrate that the finite-energy sum-rule constraint should be satisfied \cite{gauss}.  By integrating both sides of \eqref{basic_gauss} with respect to $\hat s$, it is evident that the normalization of $G_0$ corresponds to the finite-energy sum-rule.  Thus the information that is independent of the heat-evolution analysis is encapsulated in the normalized Gaussian sum-rules (NGSRs) \cite{diagonal_GSRs}
\begin{equation}
 N_0\left(\hat s,\tau\right)=\frac{G_0\left(\hat s,\tau\right)}{M_{0}\left(\tau\right)}
~,~ M_{0}(\tau)
 =\int\limits_{-\infty}^\infty 
G_0 (\hat s,\tau)\;\dif{\hat{s}} ~,
\label{N_gq}
\end{equation}
related to the spectral function via
\begin{equation}
N_0(\hat{s},\tau,s_0) = \frac{ \frac{1}{\sqrt{4\pi\tau}} \int_{t_0}^{\infty} 
   \exp\left[\frac{-(\hat{s}-t)^2}{4\tau} \right] \rho(t)\;\dif{t}}{\int_{t_0}^{\infty}
   \rho(t)\;\dif{t}} ~.
   \label{NGSR}
   \end{equation}

\begin{figure}[hbt]
\centering
\includegraphics[scale=0.6]{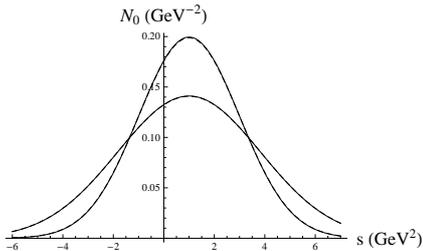}
\caption{
Comparison of the phenomenological side of the NGSR \eqref{NGSR} for a single resonance of mass $m=1.0\,{\rm GeV}$ (solid curves) 
and a square pulse 
with width $\Gamma=0.3\,{\rm GeV}$ (dashed curves).   The overlapping upper pair of curves is for $\tau=2\,{\rm GeV^4}$ and the bottom (overlapping) pair corresponds to $\tau=4\,{\rm GeV^4}$.
}
\label{width_fig}
\end{figure}

A model of two narrow resonances plus QCD continuum is used to analyze the NGSR \eqref{NGSR}.   Although superficially this would seem flawed for scalars, resonance width effects are obscured by the width $2\sqrt{\tau}$ of the Gaussian kernel  for the $\tau$ ranges where the QCD sum-rule is viable. This is illustrated in Fig.~\ref{width_fig} which compares the contributions to $N_0$ from a single resonance of mass $m$  with a square pulse  of width $\Gamma$ \cite{Elias:1998bq} centred at the same mass $m$.  It is clear from the figure that there is no discernible width effect after integration with the Gaussian kernel.   This insensitivity to resonance widths is quite distinct from Laplace sum-rules where resonance width effects influence the mass prediction that emerge from the sum-rules  in scalar channels \cite{Elias:1998bq}.

Using the two narrow resonance model, the diagonal NGSR \eqref{NGSR} becomes
\begin{gather}
\begin{split}
   {\sqrt{4\ppi\tau}}  N_0\left(\hat{s},\tau,s_0\right) & =
  r_1\exp\left[-(\hat{s}-m_1^2)^2/(4\tau)\right]
  \\&
  +       r_2 \exp\left[-(\hat{s}-m_2^2)^2/(4\tau)\right]
  \,.
\end{split}
  \label{double_N0}
\end{gather}
Since the coupling parameters satisfy 
  ${r_1+r_2=1}$, in the diagonal (glueball-glueball and quark-quark) cases, where $r_i={f_i^2}/\left({f_1^2+f_2^2}\right)$, they can be interpreted as the relative strength associated with each resonance.
  
\begin{figure}[hbt]
\centering
\includegraphics[scale=0.45]{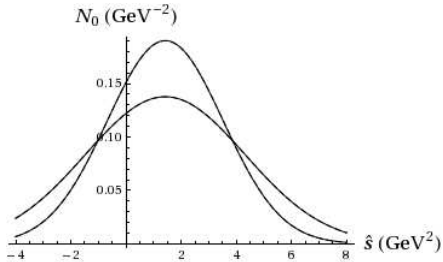}
\caption{
Comparison of the QCD
theoretical expression  for the non-diagonal NGSR $N_0^{(Gq)}\left(\hat s, \tau,s_0\right)$ with the fitted
double narrow resonance phenomenological model. The upper set of overlapping curves are for $\tau=2\,{\rm GeV^4}$ and the bottom set of overlapping curves is for $\tau=4\,{\rm GeV^4}$.
}
\label{nondiag_fit_fig}
\end{figure}

The resonance parameters and continuum are determined by fitting the model to the QCD prediction.  The QCD expressions for $N_0$ corresponding to the three correlators  (\ref{diag_glue}--\ref{non_diag}) are too lengthy to be presented here; complete expressions may be found in Refs.~\cite{Harnett:2008cw,diagonal_GSRs}.  In addition to $\Lambda_{\overline{MS}}\approx 300\,{\rm MeV}$ and the PCAC value for the quark condensate, the following QCD parameters are used in our analysis \cite{pdg,DIL,condensates}:
\begin{gather}\label{dimfour}
 \langle\alpha G^2\rangle
      = (0.07\pm 0.01)\, {\rm GeV^4} \,,
 \\
 \left\langle \bar q \sigma Gq\right\rangle=M_0^2\left\langle\bar q q\right\rangle~,~M_0^2=\left(0.8\pm 0.1\right) \,{\rm GeV^2}\,,
\label{mixed_cond_value}
\\
  n_{{c}} = 8.0\times 10^{-4}\ {\rm GeV^4}~,~\rho =
  \frac{1}{0.6}\ {\rm GeV}^{-1}\,,
  \\
  2.5\,{\rm MeV} <m_q(2\,{\rm GeV})<5.5\,{\rm MeV}~.
\label{quark_mass}
\end{gather}
The results of an independent fit to each correlator are given in Table~\ref{doubres_tab}, and shown for the non-diagonal (glueball-quark) correlator in Figure~\ref{nondiag_fit_fig}.   The agreement between the QCD prediction and the optimized resonance model is excellent, with no discrepancies that would be indicative of additional states or resonance width (or resonance shape) effects.
The uncertainties arising from the QCD input parameters are approximately $0.2\,{\rm GeV}$ for the masses, although there is a correlated effect that leads to a relatively stable splitting $m_2-m_1\approx 0.5\,{\rm GeV}$.  The uncertainties in the coupling strengths $r_i$ is approximately $0.1$.  Taking into account the uncertainties in the mass predictions there is a remarkable consistency independently arising from all three correlators, providing clear evidence for mixed states coupling to both $q\bar q$ and glueball currents, with the heavier state coupling more strongly   to glueball currents.

\begin{table}[hbt]
  \centering
  \begin{tabular}{||c|c|c|c|c|c||}
\hline\hline
Correlator 	&$m_1$ 		& $m_2$       	& $r_1$		& $r_2$		& $s_0$  		\\ \hline
$J_GJ_G$   	&  $0.98$ 	& $1.4$ 	& $0.28$ 	& $0.72$ 	& $2.30$ 		\\ \hline
$J_qJ_q$   	& $0.97$ 	& $1.4$ 	& $0.63$ 	& $0.37$ 	& $2.60$		\\ \hline
$J_GJ_q$ 	& $0.84$ 	& $1.4$ 	& $0.44$ 	& $0.56$ 	& $2.75$		\\ \hline\hline
  \end{tabular}
 \caption{Analysis results from the diagonal and non-diagonal NGSRs of gluonic and $\bar q q$ currents  in the double narrow resonance model.  Central values of the QCD input parameters have been employed and all units are in GeV to the appropriate power.}\label{doubres_tab}
\end{table}

The relative strengths $r_i$ also reveal a remarkable consistency.  From the diagonal glueball-glueball and quark-quark  results,
\begin{gather}
 r^{(GG)}_1=0.28=\frac{{f_{1G}}^2}{{f_{1G}}^2+{f_{2G}}^2}\,,~r^{(GG)}_2=0.72=\frac{{f_{2G}}^2}{{f_{1G}}^2+{f_{2G}}^2}
\nonumber
\\
r^{(qq)}_1=0.63=\frac{{f_{1q}}^2}{{f_{1q}}^2+{f_{2q}}^2}\,,~r^{(qq)}_2=0.37=\frac{{f_{2q}}^2}{{f_{1q}}^2+{f_{2q}}^2}\,,
\nonumber
\end{gather}
 one can determine two possible solutions for the non-diagonal glueball-quark case
 \begin{equation}
 r^{(Gq)}_1=\frac{f_{1G}f_{1q}}{f_{1G}f_{1q}+f_{2G}f_{2q}}=\begin{cases}+0.45\\-4.4\end{cases}
\end{equation}
where the numerical index on the  coupling refers  to the state of mass $m_i$ and $G,q$ refers to glueball or quark currents.
The positive solution is in superb agreement with the predicted value for the glueball-quark correlator (see final row of Table~\ref{doubres_tab}).
Following Ref.~\cite{lattice} the state couplings can also be used to determine an effective mixing angle $\phi$
\begin{equation}
\tan^2\phi=\left\vert
\frac{\langle 0\vert J_g \vert 1\rangle\, \langle 0\vert J_q\vert 2\rangle}{\langle 0\vert J_g\vert 2\rangle\, \langle 0\vert J_q\vert 1\rangle} \right\vert~,
\end{equation}
where $\vert 1\rangle$ and $\vert 2\rangle$ respectively correspond to the states with mass $m_1$ and $m_2$.
Taking into account the uncertainties in the Table~\ref{doubres_tab} values  leads to  $\phi=54^\circ\pm 4^\circ$ corresponding to states that couple to nearly-maximal mixtures of glueball and $q\bar q$ currents.

In summary, the three possible correlation functions of glueball and $q\bar q$ scalar currents provide signals of mixing between scalar glueballs and $q\bar q$ mesons.  For  significant mixing the non-diagonal correlator must be non-zero, and this occurs through 
chiral-violating non-perturbative effects from QCD condensates and instantons.  Gaussian sum-rules are  able to probe multiple states with equal sensitivity, ideal for studying these mixing effects.  They also show minimal sensitivity to resonance width effects, which is important in the scalar isoscalar sector where the known states have comparatively large widths.

The independent analysis of all three normalized  Gaussian sum-rules reveals a remarkably consistent scenario of two resonances with a mass splitting of $0.5\,{\rm GeV}$ coupling to nearly-maximal combinations of glueball and $q\bar q$ currents, 
with the heavier state having a slightly larger
glueball component.
This result suggests that unique signatures of pure glueball  states would be obscured, complicating definitive experimental identification of glueballs.  Our results provide complementary supporting evidence for the mixing of $q\bar q$ and gluonium to be  manifested in the scalar hadronic spectrum as a lighter state on the order of $1\,{\rm GeV}$ and a heavier state on the order of $1.5\, {\rm GeV}$ \cite{narison_review,other_stuff,lattice}.

\noindent
{\bf Acknowledgements:}  The authors  are grateful for financial support from the Natural Sciences and Engineering Research Council of Canada (NSERC).  Many thanks to Ailin Zhang and Hong-Ying Jin for helpful discussions. 













\end{document}